\documentclass[oldversion,preprint]{aa}

\usepackage{graphicx}
\usepackage{txfonts}
\usepackage{longtable}
\usepackage{multirow}

\begin{document}

\title{On the Calibration of Full-polarization 86\,GHz Global VLBI Observations}
\titlerunning{Calibration of GMVA 86\,GHz Observations}

   \author{I. Mart\'i-Vidal\inst{1,2} \and T.P. Krichbaum\inst{1} \and A. Marscher\inst{3}
            \and W. Alef\inst{1} \and A. Bertarini\inst{1,4} \and U. Bach\inst{1}
            \and 
            F.K. Schinzel\inst{1}\thanks{Now at Department of Physics and Astronomy, 
            University of New Mexico, Albuquerque, NM 87131, USA} 
            \and H. Rottmann\inst{1} \and J.M. Anderson\inst{1} 
            \and J.A. Zensus\inst{1} \and M. Bremer\inst{5} \and S. Sanchez\inst{6}
            \and M. Lindqvist\inst{2} \and A. Mujunen\inst{7} }

   \institute{
             Max-Planck-Institut f\"ur Radioastronomie (MPIfR),
             Auf dem H\"ugel 69, D-53121 Bonn (Germany)\\
             \email{ivan.marti-vidal@chalmers.se} 
             \and
             Onsala Space Observatory (Chalmers University of Technology),
             Observatoriev\"agen 90, SE-43992 Onsala (Sweden)
             \and
             Institute for Astrophysical Research, Boston University, 
             725 Commonwealth Avenue, Boston, MA-02215 (USA)
             \and
             Institute of Geodesy and Geoinformation, Bonn University,
             Nussallee 17, D-53115 Bonn (Germany)
             \and
             Institut de Radioastronomie Millim\'etrique (IRAM), 
             300 Rue de la Piscine, F-38406 Saint Martin d'H\`eres (France)
             \and
             Instituto de Radioastronom\'ia Milim\'etrica (IRAM),
             Av. Divina Pastora 7, N\'ucleo Central, E-18012 Granada (Spain)
             \and
             Mets\"ahovi Radio Observatory (Aalto University),
             Mets\"ahovintie 114, FI-02540 Kylm\"al\"a (Finland)
}


\date{Version 12. March 6, 2012.}

\abstract{
We report the development of a semi-automatic pipeline for the calibration 
of 86\,GHz full-polarization observations performed with the Global 
Millimeter-VLBI array (GMVA) and describe the calibration strategy followed 
in the data reduction. Our calibration pipeline involves non-standard 
procedures, since VLBI polarimetry at frequencies above 43\,GHz is not yet 
well established. We also present, 
for the first time, a full-polarization global-VLBI 
image at 86\,GHz (source 3C\,345), as an example of the final product 
of our calibration pipeline, and discuss the effect of instrumental 
limitations on the fidelity of the polarization images. Our calibration
strategy is not exclusive for the GMVA, and could be applied on other 
VLBI arrays at millimeter wavelengths. The use of this pipeline will allow 
GMVA observers to get fully-calibrated datasets shortly 
after the data correlation. 
}


\maketitle

\section{Introduction}

The Global Millimeter-VLBI array (GMVA) is the result of a collaboration
of a group of radio observatories, led by the Max-Planck-Institut f\"ur 
Radioastronomie (MPIfR), interested in performing 
astronomical VLBI observations at millimeter 
wavelengths\footnote{See \texttt{http://www.mpifr-bonn.mpg.de/div/vlbi/globalmm}}. 
Currently, the GMVA is formed by the radio telescopes at Effelsberg (100\,m, MPIfR, 
Germany), Pico Veleta (30\,m, IRAM, Spain), Plateau de Bure (six 15\,m antennas 
working in phased-array mode, IRAM, France), Onsala (20\,m, Sweden), Mets\"ahovi 
(14\,m, Finland), and a subset of the Very Long Baseline Array\footnote{The VLBA 
(National Radio Astronomy Observatory, NRAO) comprises ten identical antennas of 
25\,m diameter, spread across the USA.} (i.e., all the VLBA antennas equipped with 86\,GHz 
receivers, which are those at Brewster, Owens Valley, Mauna Kea, Pie Town, Kitt Peak, 
Fort Davis, Los Alamos, and North Liberty). Some technical details of these 
antennas are given in Table \ref{Techs}. It is planned that additional antennas
(e.g., the 40\,m telescope at Yebes Observatory, Spain; the NRAO 100\,m Green Bank 
Telescope, USA; the 50\,m Large Millimeter Telescope, LMT, Mexico; the 64\,m 
Sardinia Radio Telescope, SRT, Italy; and, later, the Atacama Large mm-submm Array, ALMA, Chile) 
will join the global 3mm-VLBI effort in the future.
Owing to its large number of participating telescopes and a
coordinated observing strategy, based on efficient observing time allocation, the GMVA 
is capable of providing good-quality images with a high spatial resolution ($40-70 \mu$as) 
at 86 GHz. On its most sensitive baselines (i.e., to the IRAM and Effelsberg telescopes), the 
GMVA offers a $\sim3-4$ times higher sensitivity and a $\sim 2$ times higher angular 
resolution than the stand-alone VLBA. This makes it possible to obtain detailed high angular
resolution and high quality 
images of emission regions which appear self-absorbed (and are therefore 
invisible) at lower frequencies. The very high spatial resolution achievable with 
the GMVA is crucial for the understanding of high-energy 
astrophysical phenomena (e.g. physical processes in Active Galactic Nuclei, AGN, 
and in the vicinity of supermassive black holes).

\begin{table*}
\caption{Technical details of the GMVA stations.} 
\label{Techs} 
\centering 
\begin{tabular}{c c c c c c} 
\hline\hline 
Name &      Diameter & $T_{\mathrm{sys}}$ at Zenith & Calib & SEFD & Pol. leakage (May 2010) \\ 
     &         (m)   &   (K)                        &       & (Jy) &            (\%)       \\
\hline 
Effelsberg (EF)      & 100  & 130 &  {\em Diode}    &   929 & $7\pm3$ \\
Plateau de Bure (PB) & 34.8 & 90  &  {\em Average}  &   409 & $3\pm3$  \\
Pico Veleta     (PV) & 30   & 90  &  {\em Chopper}  &   643 & $2\pm2$ \\
VLBA                 & 25   & 100 &  {\em Diode}    &  2941 & $6\pm3$  \\
Onsala     (ON)      & 20   & 300 &  {\em Chopper}  &  6122 &      --     \\
Mets\"ahovi   (MH)   & 14   & 300 &  {\em Diode}    & 17647 & $4\pm2$ \\
\hline 
\end{tabular}
\tablefoot{Values for PB are given for the combined array (i.e., in phased-array mode). 
{\em Diode} stands 
for the common {\em on-off} noise-diode method (or equivalent), which does not correct for the 
atmospheric opacity; {\em Chopper} stands for the {\em hot-cold} chopper-wheel method (or equivalent; 
e.g. Penzias \& Burrus \cite{Penzias}), which corrects for the opacity; and {\em Average} stands for the 
$T_{\mathrm{sys}}$ average of all the PB antennas (accounting for the phasing efficiency and applying 
model-based estimates of the opacity). SEFD is the system equivalent flux density. The polarisation 
leakage column (D-term) is the average amplitude of the 
polarization-leakage factors, as estimated from the fitting for all the sources observed in the 
GMVA session reported in this paper (see Sect. \ref{V}). Onsala only records LCP (hence, there are no 
D-terms estimated).}
\end{table*}

In this paper we describe the steps required for the calibration of GMVA observations,
from the subband phase calibration (i.e., the alignment of phases and delays in the different 
observing subbands) to the global fringe fitting (GFF; Schwab \& Cotton 
\cite{Schwab}) and the polarization calibration. 
This paper focuses only on the technical aspects in the data calibration and reduction,
which at millimeter wavelengths deviates in some details from the standard data analysis
(which is typically applied at longer cm-wavelengths).
The scientific exploitation of the data is of course a matter of the principal 
investigators (PIs) of the projects approved for GMVA observations. 

A special motivation for this paper is the fact that VLBI polarimetry at frequencies 
above 43\,GHz is not standard, nor well established. Extending the frequency coverage of 
VLBI polarimetry to higher frequencies is important for a better understanding of the 
details in jet physics (e.g. Homan et al. \cite{Homan2009}; O'Sullivan et al \cite{Sullivan2011};
G\'omez et al. \cite{Gomez2011}) or the origin and 
launching mechanisms of jets near the central black hole in AGN 
(e.g., Broderick \& Loeb \cite{Broderick2009}; Tchekhovskoy et al. \cite{Tchek2011}; Mc\,Kinney 
et al. \cite{McKinney2012}). We therefore believe that it is important to discuss the 
possibilities and limitations of polarimetric VLBI observations at mm-wavelengths, which so 
far are not yet fully exploited. There are only a few published 86\,GHz polarization images 
(e.g. Attridge \cite{Attridge2001}; Attridge et al. \cite{Attridge2005}; G\'omez et al. 
\cite{Gomez2011}), which were made using the VLBA only, and not the global, and more 
sensitive, 3mm-VLBI array.


We present as an example of our calibration strategy results obtained from part of the 
full-polarization observations taken in the GMVA session in May 2010. We also present some 
representative images (in total intensity and polarization) of the quasar 3C\,345 (one of the sources 
observed in that session). In Sect. \ref{II}, we summarize the technical 
details of the observations, and in Sects. \ref{III} to \ref{V} we depict 
the calibration strategy in chronological order: the whole phase calibration is 
described in Sect. \ref{III}; the amplitude calibration is described in Sect \ref{IV}; 
and the correction for the polarization leakage at the receivers is described in 
Sect. \ref{V}. Finally, we present sample images of 3C\,345 in Sect. \ref{VI} and 
summarize our work in Sect. \ref{VII}.

\section{Observing with the GMVA}
\label{Ib}

For logistical reasons, the GMVA observations are performed in 4--6 day-long sessions
twice per year (in spring and autumn) and the Call for Proposals shares the deadlines 
with those of the VLBA (i.e., February 1st and August 1st each year). The proposals 
are refereed individually by the participating institutes, and the ratings are then 
combined to determine what projects shall be observed.

For each observing session, the experiments belonging to different principal 
investigators (PIs) are combined in a single VLBI observing time block at all telescopes. 
Within this block time, the detailed observing schedule may be sub-divided 
in different {\em scheduling blocks} arranged to minimize the idle times 
of the telescopes and to maximize the uv-coverage for the 
observed radio sources within the given time constraints. Hence, when a source is not visible 
to the whole interferometer (because of the different rise and setting times between the USA 
and Europe), the scheduling strategy includes sub-arraying (i.e., division of the whole 
GMVA into two or more independent arrays). The use of subarrays allows the schedulers 
to optimize on-source integration times and antenna elevations, but also causes some 
difficulties in the data calibration and reduction. For instance, it is not always possible to 
assign a common reference antenna for the global fringe fitting (GFF). Hence, the 
calibration of phase-like quantities (phases, delays, and 
rates) requires a continuous re-referencing between the subarrays, which may often change 
during the GMVA session. 

Nevertheless, all the peculiarities in the data calibration due to the complex structure of 
GMVA schedules should not represent any problem for the PIs, since the bulk of the data 
calibration and editing could be performed at the VLBI correlator and data analysis center 
(e.g at the MPIfR), following the steps described in this paper.

\section{GMVA observations on May 2010}
\label{II}

This paper concerns, as a test dataset, the GMVA observations conducted between the 
6th and the 11th of May 2010. Most of the 
observing session was performed in dual-polarization mode (i.e., the left circular 
polarization, LCP, and the right circular polarization, RCP, were simultaneously observed) 
at a frequency of 86\,GHz, with an overall recording rate of 512\,Mb\,s$^{-1}$, 2-bit 
sampling in Mark5B format. Four 16\,MHz subbands were used at each polarization. For each subband, the 
correlator produced 32 spectral channels. 
In the correlation process, all possible combinations of the polarizations were correlated,
to yield all four Stokes products.

The full set of GMVA antennas participated in these observations.
The observations were divided in scans of $\sim$7\,minutes. There were a total of 18
AGN observed with different overall on-source times.

\section{Phase Calibration}
\label{III}

The phase calibration is the most critical and time-consuming part in the data reduction, 
especially at 86\,GHz, because of strong atmospheric and instrumental phase instabilities. 
The full process of 
phase calibration (with the exception of the eventual phase self-calibration involved 
in the source imaging) was performed using the NRAO Astronomical Image Processing System 
(AIPS). We used AIPS in batch mode by writing several scripts in ParselTongue (a Python 
interface to AIPS; see Kettenis et al. \cite{Kettenis2006}). This process involves 
the following main steps.

\begin{itemize}

\item {\em Preliminary calibration}. We corrected the effect of the changing parallactic 
angle of each antenna. The effects of the Nasmyth 
mount of the Pico Veleta station were also corrected (see Dodson \cite{Dodson}).

\item {\em Subband phase calibration}. The independent oscillators of the single-sideband 
mixers introduce unknown phase offsets in each subband. In addition, due to the different 
lengths in the signal paths, there may be slightly different delays and phases among the 
subbands at each station. These delays and phases were referred to one 
(reference) antenna.

\item {\em Global fringe-fitting on the multi-band data}. We found the antenna-dependent 
multi-band gains (i.e., delays, phases, and phase rates, over the whole band) in all 
the observations.

\item {\em Polarization calibration}. We found the 
delay and phase difference between subbands in the cross-hand (i.e., RL and LR) correlations.

\end{itemize}

We emphasize that the long duration of the GMVA observing sessions ($\sim 3 - 5$ days) and the 
subarraying may affect the results of each step in the calibration, as we discuss in the 
following subsections.

\subsection{Subband phase calibration}
\label{BWSynth}

A common strategy for the correction of the different delays and phases of the 
subbands is to use the so-called {\em phase-cal injection
tones}, which are sharp pulses injected in the signal path, close to the receiver
horn. However, 
this approach is not possible for the GMVA, since the 86\,GHz receivers of the
VLBA do not have phase-cal injection tones. In addition, the phase-cal tones at the 
European telescopes are not injected at the receiver front-end, but at a later stage in the 
signal path. Hence, there may be instrumental phase variations in the signal that cannot be 
corrected from the phase-cal tones; these variations can only be removed via the 
alternative {\em manual phase-calibration} approach. 

With a manual phase-calibration, the unknown delays and phases among subbands are estimated from 
the application of the global fringe-fitting algorithm to a set of visibilities from a bright source. Independent
solutions for the delay and phase of each subband, antenna, and polarization are 
found from the observations. Then, the delay and phase solutions for that particular subset
of visibilities are extrapolated to the whole dataset. The antenna-dependent phase solutions 
computed with the global fringe-fitting algorithm must be referred to a so-called {\em reference} antenna, which 
has assigned, by definition, a zero phase (and delay) gain.

However, the manual phase-calibration may lead to an imperfect alignment
of the phases among the subbands, mainly due to possible drifts in the electronics of the 
receiving systems during the relatively long duration of a GMVA session. Moreover, the 
many subarraying conditions present in the GMVA observations make it
impossible to assign the {\em same} reference antenna to the whole dataset. In addition, it 
is difficult to find scans of bright sources simultaneously observed with the whole 
interferometer, since weather or station-related problems may cause missed 
calibrator observations; furthermore, the calibrators are quite 
variable at 86\,GHz, so it is not easy to select the best calibrator sources at the time of 
the schedule preparation. 
Our script for the calibration of GMVA observations overcomes these drawbacks of 
manual phase-calibration in the following way.

\begin{enumerate}

\item The script performs a global fringe fitting (using the AIPS task FRING) to the whole 
set of observations. It finds 
independent solutions for the phases and delays of each subband and polarization. Different 
reference antennas may be used by FRING if the main reference antenna (e.g., Los Alamos) 
is missing in a particular subarray and/or time. We notice, though, that any change 
of reference antenna made by FRING does not affect our final calibration (see below).

\item From all the FRING solutions, the script filters only those with the highest 
signal-to-noise ratio (SNR). A typical cut-off is SNR$\geq 20$. 

\item The remaining solutions are arranged by antenna and reference antenna,  
and the delays and phases are referred to those of a given (reference) subband and 
polarization. (i.e., the phase and delay differences between subbands are calculated, 
in order to remove these purely instrumental contributions from the data). 

\item The resulting delay and phase differences of each antenna are binned using a 
median-window filter (MWF) and the bins are linearly interpolated in time. Different averaging 
and interpolation schemes may be applied and visually checked, until a satisfactory 
time interpolation of the phases and delays at all the antennas is obtained.

\item The interpolated phases and delays are applied to the whole dataset and 
re-referenced, when necessary, to the main (i.e., the most commonly appearing) 
reference antenna. 

\end{enumerate}

We emphasize that even if the main reference antenna is not present in a particular subarray 
and time, it is still possible to re-reference the delays and phases to that antenna, by 
means of bootstrapping 
(i.e., from a phase connection through the interpolated solutions of all the antennas).
We notice further that this 
algorithm is applied transparently and homogeneously to the different subarrays in the data,
and in such a way that the phase gain of the main reference antenna is always zero. 

\begin{figure*}
\centering
\includegraphics[width=18cm]{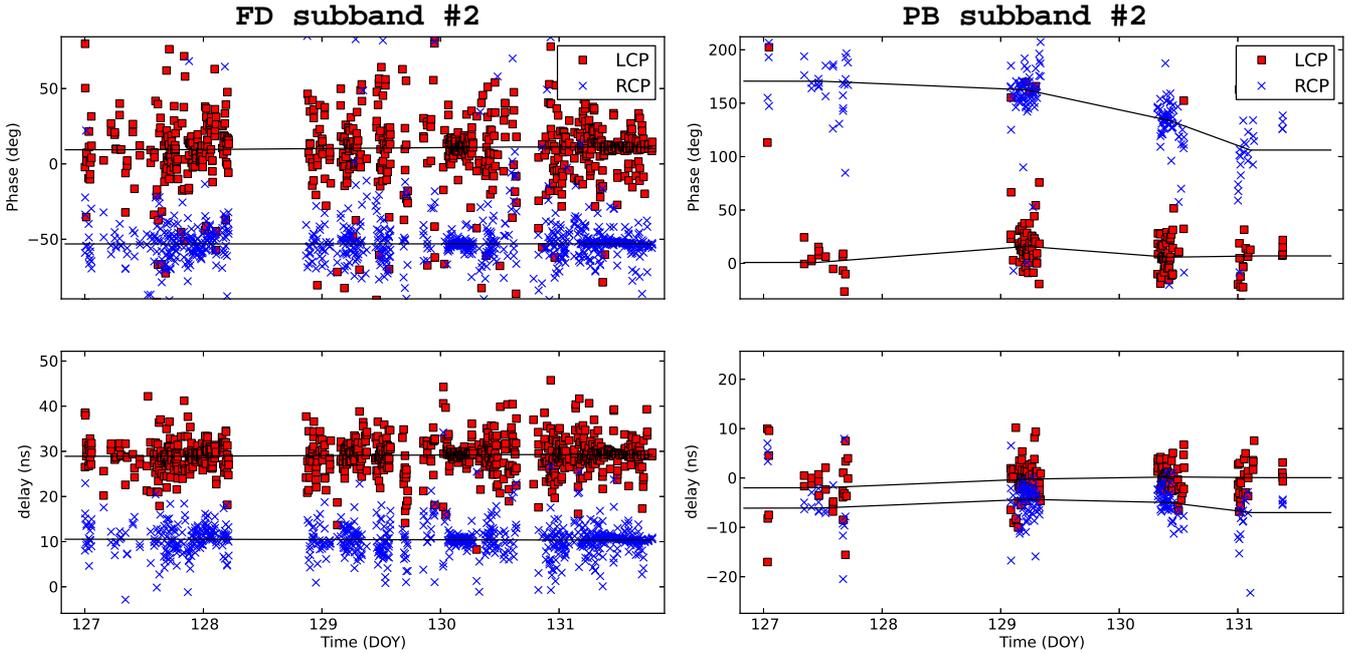}
\caption{Single-band phases (upper figures) and delays (lower figures) in the second subband
of Fort Davis (left) and Plateau de Bure (right), referred to Los Alamos, for both 
circular polarizations. Times are given in day of the year (DOY). The gains are referred 
to the first subband in the 
LCP polarization. Solid lines are the interpolations applied to 
calibrate the data. Notice that these plots show data from {\em all} sources.}
\label{IF-aligner.eps}
\end{figure*}

In addition, this approach does not assume that the phase and delay differences between subbands 
remain constant during the whole session, but allows us to check any drift in the electronics
of the receivers. We show in Fig. \ref{IF-aligner.eps} some representative plots generated with
our script. The subband used as reference was the one corresponding to LCP and 
lowest sideband frequency. For the case of Fort Davis (antenna code FD), it can be seen that 
the phases (with an average of $10.3\pm0.9$\,deg. for LCP and $-53.1\pm0.3$\,deg. for RCP) and 
delays (with an average of $29.09\pm0.18$\,ns for LCP and $10.45\pm0.10$\,ns for RCP) remain 
remarkably constant, as it is also the case for most of the antennas. However, 
a drift is seen in the phases of the second subband of Plateau de Bure (antenna code PB) for the RCP 
polarization (average phase of $6.07\pm5.01$\,deg. for LCP; $124.4\pm68.6$\,deg. for RCP). 
The overall drift in this subband is larger than 50\,deg. through the whole session, which translates 
into an average drift of $\sim 10$\,deg. per day. 
Since PB is a phased array, the drift seen in Fig. \ref{IF-aligner.eps} may be due to the 
signal pre-processing before arriving at the recording system, although similar drifts have been 
found at other antennas (e.g., Effelsberg) during the on-going analysis of other GMVA 
observations (not reported here). We are currently analyzing in 
deeper detail what could be the reason of these unexpected phase drifts between subbands.

\subsection{Fringe fitting}
\label{GFF}

Once the manual phase calibration has been performed to yield higher sensitivity, it is 
possible to combine the data from all the subbands and estimate the (multi-band) phases, 
delays, and rates for each antenna, source, and time. This step is the so-called 
{\em multi-band fringe fitting}. 

Since we are usually dealing with weak sources (in terms of the sensitivity of the antennas) 
and the coherence time of the signals is relatively short (due to the effect of rapidly-changing 
atmosphere at high frequencies) it is not straightforward to select the best combination of 
parameters for the fringe-fitting algorithm.
On the one hand, a short coherence time calls for short integration times, to avoid 
a reduction of amplitude. This is so because non-linear drifts in the phase would 
degrade the fringes, thus broadening (or even smearing out) the 
fringes. On the other hand, a short integration time reduces the 
chance of a successful fringe detection.

\begin{figure}
\centering
\includegraphics[width=9cm]{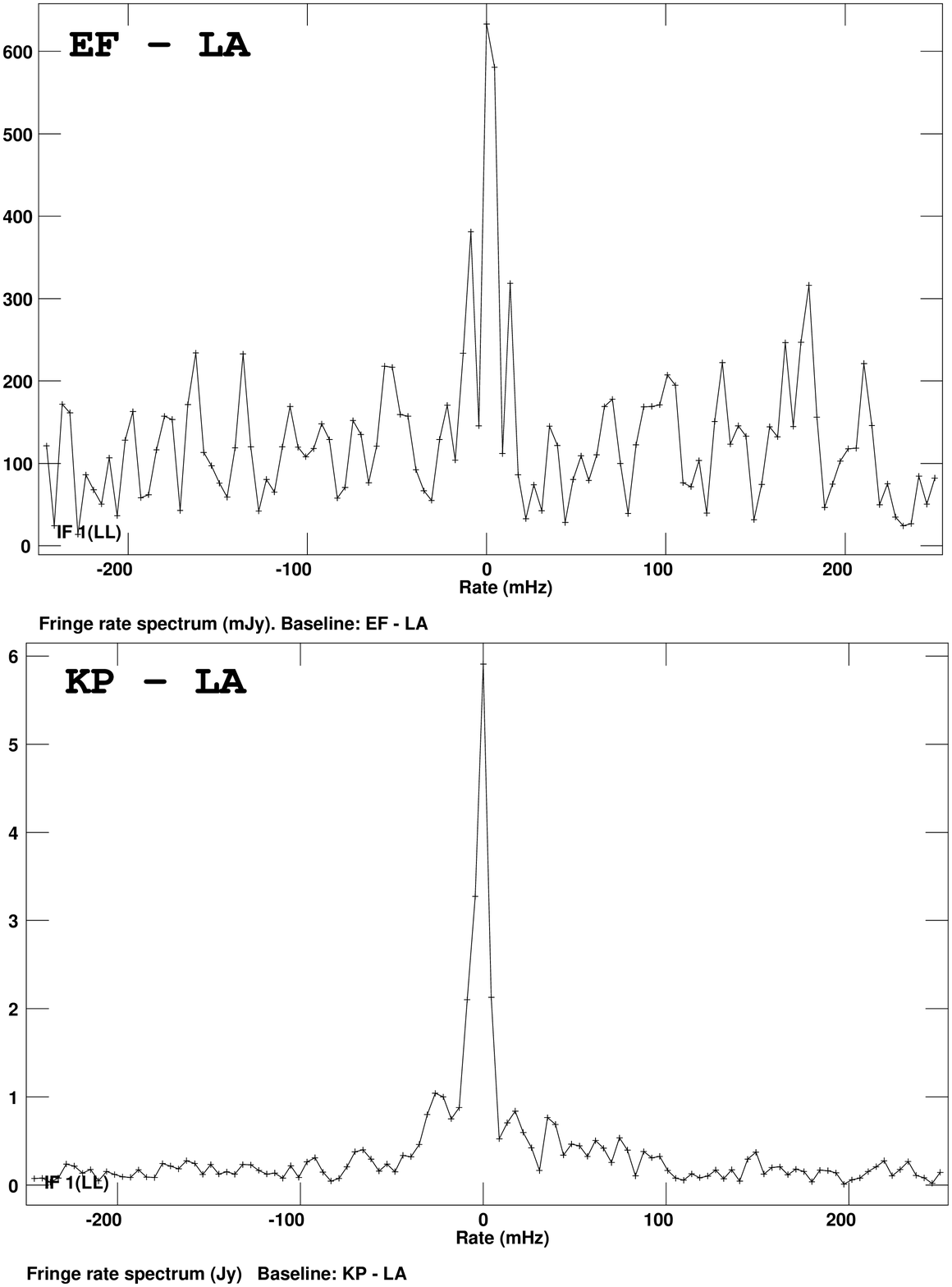}
\caption{Fringe-rate spectra at the baselines of Effelsberg to Los Alamos (EF-LA, baseline of 
7831\,km) and Kitt Peak
to Los Alamos (KP-LA, baseline of 752\,km), for an observation of source 3C\,273B with an 
integration time of 4 minutes.}
\label{Rate}
\end{figure}

We estimated the best integration time to be used on GMVA data by analyzing the performance of 
the global fringe-fitting 
for different integration times. For an integration time of 3--4 minutes, the number of good 
(i.e., high SNR) solutions is maximized with respect to bad, or failed, ones. Since an integration 
time of 3--4 minutes is much longer than the actual (expected) atmosphere coherence time at 86\,GHz 
(i.e, $\sim 10-20$ seconds), our results indicate that the changes in the fringe rate due to the atmosphere 
are not so severe and/or systematic as to break down the phase coherence during an integration time 
longer than the expected $\sim 10-20$ seconds, although the exact coherence time will depend, 
of course, on the weather conditions at each station (humidity, wind speed, etc.). In other words, 
the phase fluctuations are mostly around an average slope (on a time scale much longer than that of 
the wrapping of the phase, for reasonably good weather conditions). Hence, if we apply 
the global fringe fitting using long integration times, we will be able to estimate and remove the main slope in 
the time evolution of the visibility phases, thus improving the signal coherence (see, e.g., 
Rogers et al. \cite{Rogers1984}; Baath et al. \cite{Baath1992}; and Rogers, Doeleman, \& Moran 
\cite{Rogers1995}, for additional discussions on the phase coherence in high-frequency VLBI 
observations).
As an example of the quality in the coherence of the 
GMVA phases, Fig. \ref{Rate} shows the fringe-rate spectra at two baselines (Effelsberg to Los Alamos
and Kitt Peak to Los Alamos) for an observation of source 3C\,273B, with an integration time 
of 4 minutes. Notice the sharp peaks in the fringe rates after such a long integration time (and 
especially for EF-LA, which is one of the longer baselines).

\begin{figure}
\centering
\includegraphics[width=9cm]{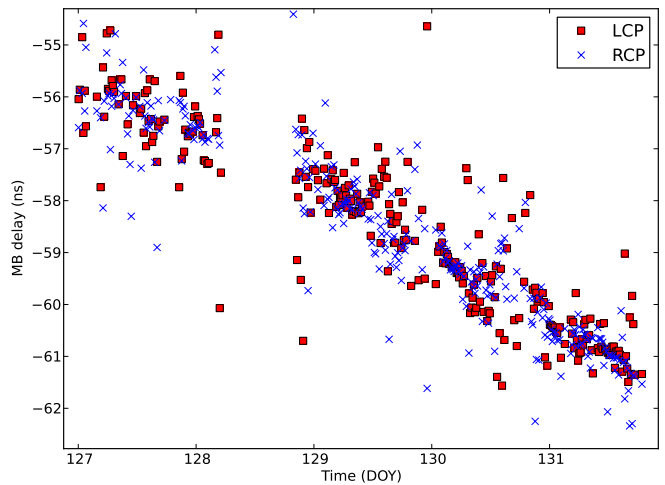}
\caption{Multi-band delay between the antennas at Brewster and Los Alamos vs. time. 
Notice that data of {\em all} sources have been included in this plot.}
\label{MBDelay}
\end{figure}

Based on these results, our script for the GMVA calibration uses a mixed approach, to optimize the 
performance of the global fringe fitting. 
First, a preliminary fringe fitting is executed using a long integration time (4 minutes) and a low SNR cut-off 
(SNR$>$4.5). Then, the script reads the estimated antenna delays and bins them in time using a median window filter. 
Finally, the script averages and interpolates the bins and applies the interpolated delays to the whole dataset. 
We notice that the bulk of the multi-band delays is expected to depend only on the antennas, and 
be almost independent of source structure. Hence, it is reasonable to homogeneously apply the delays 
interpolated by our script to the visibilities of all the observed sources.
Indeed, given that we perform the manual phase calibration by referring the phases and delays of 
one polarization to those of the other, it is also expected that the remaining multi-band delays 
will be very similar for the LCP and the RCP data. We show in Fig. \ref{MBDelay}, the 
multi-band delays computed by FRING for the baseline of Brewster to Los Alamos. The figure 
clearly shows how the delays are very similar for the different sources, and for both polarizations,
through the whole GMVA session. 

This initial estimate of the antenna delays allows us to perform a second fringe fitting with a 
shorter solution interval ($\sim$2 minutes), but using a much narrower window for the delay search 
(1--10\,ns, depending on the scatter in the delay solutions from the first fringe-fitting run). 
Since the integration time in the second fringe-fitting run is shorter, the resulting rates and phases 
will follow better the behavior of the rapidly-changing atmosphere over each antenna.

The script then tries to recover fringes that could not be 
found in the previous runs of FRING. It finds out all the combinations of antennas, sources, 
and times where there were no FRING solutions. Then, these data are pre-calibrated, using a 
linear interpolation of the nearby good FRING solutions, and a new iteration of FRING is run.
However, this time, only the failed antennas of each scan are included in the fit. This 
approach could be understood as a robust iteration in the fringe search, and minimizes the 
number of discarded (i.e., edited) visibilities, although a lower SNR cut-off ($\sim 3.5$) is 
necessary to decrease the number of failed solutions (by 15--20\%). 
The overall amount of visibilities lost because of a non-detection of fringes is $\sim$10--20\%, 
for an SNR cut-off of 4--5 (for this particular dataset).

As a final step, the script corrects for the effect of the slightly different rates and delays found 
by FRING between the RR and LL correlations. Even a small rate difference of just a few mHz between
the RR and LL correlations may translate, after the calibration, into undesired drifts in 
the RL and LR phases between subbands, thus making it very difficult to later perform a 
reliable correction of the instrumental polarization (see next section).

\subsection{Polarization calibration}

At this stage in the data calibration, all the subband phases and delays in the parallel-hand 
visibilities (i.e., the LL and RR correlations) are aligned, as described in Sect. 
\ref{BWSynth}, and the residual multi-band delays, phases, and rates are fitted and calibrated 
out, as described in the previous section. Now the only remaining instrumental effects in the 
data are due to the instrumental polarization. On the one hand, there are still delay and 
phase differences between the subbands of the the cross-hand correlations; on the other hand, 
there is a polarization leakage in the receivers of the antennas that must be estimated and 
corrected. 
Calibration of the polarization leakage is described in Sect. \ref{V}. 
In the present section, we describe the procedure used to calibrate the remaining delay and 
phase differences between the subbands of the cross-hand (i.e., RL and LR) correlations.

Any difference between the path of the RCP and LCP signals at the main reference antenna 
(i.e, the antenna with null phase gains after the manual phase calibration described in Sect. 
\ref{BWSynth}) maps 
into a phase and delay difference in the subbands of the cross-hand correlations. With the lack
of useful phase-cal tones, this 
difference can only be corrected if the cross-hand correlations are fringe-fitted.  
Our script for the calibration of GMVA observations makes use of the AIPS task BLAVG to 
average the cross-hand correlations of all the baselines related to the reference antenna, 
and exports them to a separate file. Then, the script runs FRING on the visibilities contained 
in that file and the AIPS 
task POLSN is applied to the FRING output (POLSN just refers all the solutions of one 
polarization to the other, and applies the resulting gains to all antennas.) Finally, the 
script filters out the gains of scans with a low SNR (lower cut-off of SNR$\leq$7), bins the 
remaining gains using a median window filter, and interpolates between bins.

\begin{figure*}
\centering
\includegraphics[width=18cm]{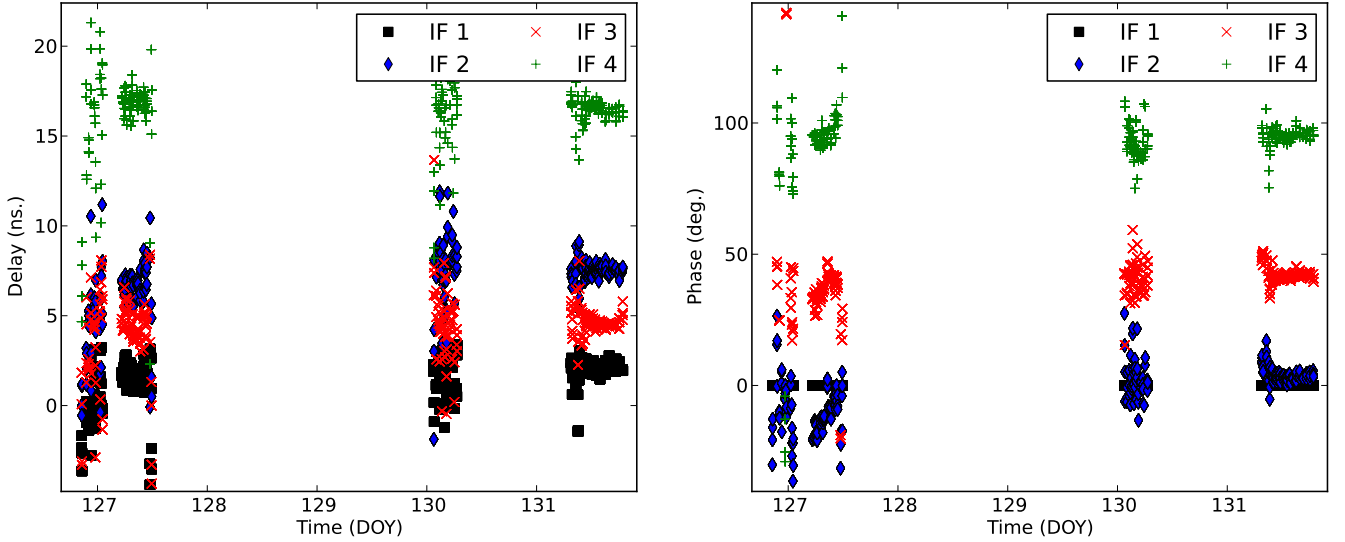}
\caption{Left, delay differences between LCP and RCP at Los Alamos. Right, phase differences 
at the same station (referred to those in the subband at the lowest frequency, IF1).}
\label{Crosshands}
\end{figure*}

We show in Fig. \ref{Crosshands} the cross-hand phases and delays found by FRING (and 
re-referenced by POLSN) for the dataset reported here. These quantities are stable 
during the whole observing session, although we notice that such a stability is found 
as long as the rates applied for the calibration of the LL and RR correlations (i.e., 
the rates found by the global fringe fitting as described in Sect. \ref{GFF}) are set to be {\em always} 
equal\footnote{In the practice, our script applies the weighted average of the RR and 
LL rates to both polarizations.}. It is indeed expected that the rates only depend on 
source coordinates, station position, and clock models, being thus equal for both 
polarizations.
If we were to calibrate the data using the rates just estimated by the fringe fitting, the small
differences that might appear between the RR and LL correlations (because of the noise 
effect in the fringe search) would introduce phase-drifts and delay differences into the 
signals of the RCP and LCP subbands of the main reference antenna, thus preventing the 
polarization calibration. 

\section{Amplitude calibration}
\label{IV}

At high radio frequencies, the atmospheric absorption becomes significant (signal attenuations 
of 10 or 30\% are not uncommon at 86\,GHz). Hence, the atmospheric opacity must be taken into 
account in the amplitude calibration of the GMVA data. The atmospheric opacity $\tau$ is 
estimated at each station (and for each time) using the well-known formula

\begin{equation}
\tau = \log{\left(1-\frac{ T_{\mathrm{sys}} - T_{\mathrm{rec}} }{T_{\mathrm{amb}}}\right)},
\label{Opac1}
\end{equation}

\noindent where $T_{\mathrm{sys}}$ is the (opacity-uncorrected) system temperature and 
$T_{\mathrm{rec}}$ is the temperature of the receiver. In this equation, it is assumed that 
the sky temperature (i.e., $T_{\mathrm{sys}} - T_{\mathrm{rec}}$) 
is equal to the average temperature of the atmosphere ($T_{\mathrm{amb}}$) corrected by the 
absorption factor $\exp{(\tau)}$. The spill-over correction and the antenna temperature due to the 
source are very small quantities (less than a few K), so that they can be neglected. If a noise diode 
is used for the signal calibration, $T_{\mathrm{sys}}$ is 
directly measured at the backend of each antenna receiver and for each scan; 
$T_{\mathrm{amb}}$ can be estimated from the weather monitoring at each station. If the 
calibration strategy is based on a chopper wheel, the system directly measures the 
opacity-corrected system temperature (i.e., $T_{\mathrm{sys}}\exp{(\tau)}$). In regard to 
$T_{\mathrm{rec}}$, it is assumed to be a stable quantity at a time scale of one day or more.

\begin{figure*}
\centering
\includegraphics[width=18cm]{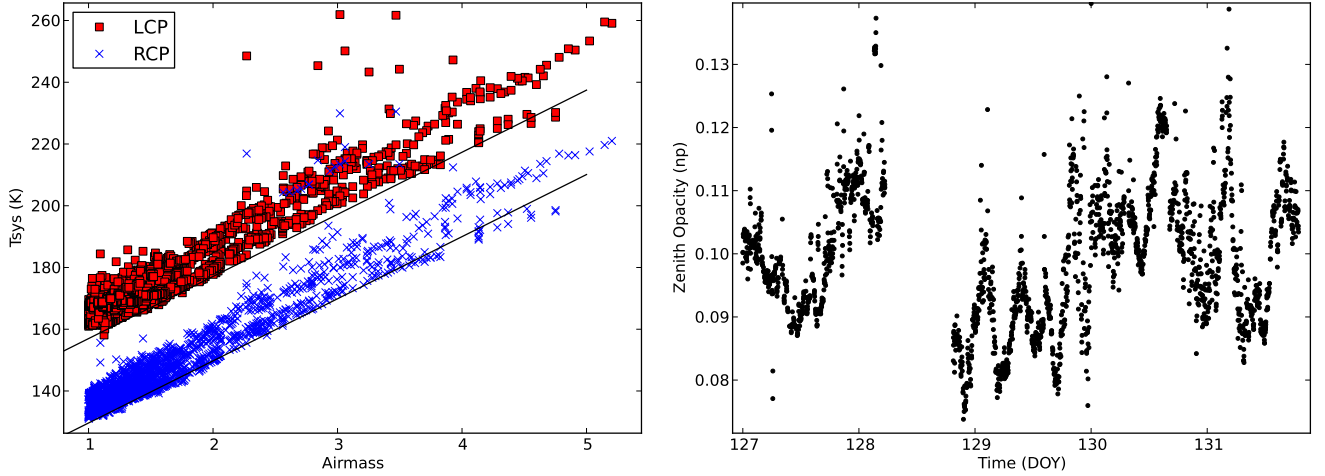}
\caption{Left, system temperatures ($T_{\mathrm{sys}}$) measured at Los Alamos vs. airmass. Straight
lines are fits of linear models to the lower envelopes of the $T_{\mathrm{sys}}$ distributions. The 
receiver temperature is estimated as the extrapolation of the lower envelope to a null airmass.
Right, zenith opacity (i.e., opacity multiplied by the sine of the antenna elevation) at Los 
Alamos vs. time, as estimated using Eq. \ref{Opac1}.}
\label{TrecFig}
\end{figure*}

We estimate the receiver temperature for each station (and polarization) by fitting the lower envelope 
of the  $T_{\mathrm{sys}}$ vs. airmass distribution with a linear model. Then, the extrapolation of that 
model to a null airmass gives us a good estimate of the receiver temperature in the time range 
considered. We show in Fig. \ref{TrecFig} (left) a sample plot of $T_{\mathrm{sys}}$ vs. airmass, together with
the fit to the lower envelope, for the Los Alamos station. It can be seen in the figure that 
the receiver temperature is slightly different for each polarization. However, we notice that our script 
forces the slopes of the lower envelopes to be the same at both polarizations 
(since the contribution of the atmosphere to $T_{\mathrm{sys}}$ is independent of the polarization). We show 
in Fig. \ref{TrecFig} (right) the resulting time evolution of the opacity at zenith (i.e., corrected by the 
sine of the elevation) at Los Alamos. Once the opacity is known, the corrected system temperature, $T^{*}$, is 
easily computed as

\begin{equation}
T^{*}_{\mathrm{sys}} = T_{\mathrm{sys}}\exp{(\tau)}.
\label{Opac2}
\end{equation}

There may be cases where an opacity correction has been already applied to the 
$T_{\mathrm{sys}}$ values provided by a given station (e.g., Plateau de Bure, Pico Veleta, or Onsala). In 
those cases, a {\em differential} (or refined) opacity correction 
can still be applied using the approach described here\footnote{At IRAM opacities are obtained using
an atmospheric model, so are not free from assumptions.}.
There is also the possibility of applying a constant
(zenith-)opacity correction to those antennae where it is not possible to find precise estimates of the 
receiver temperature. In any case, our goal in the processing of each GMVA dataset is to provide the 
end user with a calibration table including all the opacity-corrected gains, as well as a set of AIPS-friendly 
files with all the $T^{*}_{\mathrm{sys}}$ estimates (i.e., the opacity-corrected temperatures) and the 
original $T_{\mathrm{sys}}$ values measured at each station (in case that the user would like to apply 
a different approach to correct for the atmospheric opacity).

\section{Polarization leakage}
\label{V}

The LCP and RCP signals from the sources are separated in the frontend of the receivers, and follow 
different paths in the electronics. However, the receivers are not perfect, and there is a certain level
of {\em cross-talk} between the RCP and LCP signals. Hence, the 
RCP (LCP) signal recorded at each station is indeed equal to the true RCP (LCP) signal from the source, 
plus an unknown fraction of LCP (RCP) signal modified by a phase gain. The (complex) factors that account 
for the fraction of LCP (RCP) source signal transferred to the recorded RCP (LCP) signals are the 
so-called {\em D-terms}, and may be different at each antenna and for each polarization (for a deep 
discussion on the polarization leakage and its correction with the {\em D-term} approach, see 
Lepp\"anen et al. \cite{Leppanen}). The antenna D-terms are expected to depend only on the station 
hardware, and be stable quantities over periods of the order of one year (G\'omez et al. 
\cite{Gomez2002}), although this may depend on the observing frequency. In this section, we describe 
how the D-terms are estimated, and how are their effects corrected in the GMVA observations.

Once the data are calibrated in both phase and amplitude, we perform 
a deep hybrid imaging with the program {\sc Difmap} (Shepherd et al. \cite{Shepherd}), by applying phase 
and amplitude self-calibration under reasonable limits, and taking special care with noisier data 
(since, in those cases, there may be a large probability of generating spurious components in the 
source structures; e.g., Mart\'i-Vidal \& Marcaide \cite{MartiSpurious}). To ensure optimum results
all the hybrid imaging is performed without scripting.

The final images and calibrated data are then read back into AIPS by the script, source by 
source. The AIPS task CALIB is executed to perform the correction of any possible RCP-to-LCP amplitude 
bias at the antennas (by assuming
a zero circular polarization for all the sources). The CLEAN components corresponding to the main features 
in the structure of each source are then joined with the AIPS task CCEDT (the regions defining 
the main features in the source structures have already been selected manually, after the 
hybrid imaging). 
Finally, the task LPCAL estimates the D-terms of each antenna, as well
as the polarization of the different source components.

We notice that the accuracy of the D-term determination depends on the strength of the detected
cross-polarized signal, which may be higher if the source is strongly polarized or
if the antenna has strong intrinsic cross-polarization (but it should be below of 10\%, to
avoid problems with the linear approximation used in LPCAL). The accuracy also
depends on the uniformity and range of the parallactic angle coverage and, to a lesser
extend, also on the complexity of the polarized source sub-structure.

\subsection{D-terms and image fidelity}
\label{Fidelity}

Since the D-terms are estimated using the data of each source separately, we have as 
many estimates of antenna D-terms as sources (we are currently working on the possibility 
of fitting one single set of D-terms to the visibilities of all sources, simultaneously, 
which would result in a more robust modeling of the polarization leakage). 
The final D-terms that we apply to each antenna are a weighted average of 
all estimated D-terms. Prior to 
the average, any clear outliers are removed, and the relative weights are adjusted as 
a function of the source flux density (and its fraction of polarized emission); 
the higher the signal of the source in the cross-hand correlations, the higher the 
weight of the corresponding D-terms in the average.  
This approach is very similar to 
those reported in previous publications discussing high-frequency VLBI polarimetry 
(e.g. Marscher et al. \cite{MarscherPol}).

We give the average D-term amplitudes of all the antennas in Table 
\ref{Techs} (Col. 6). 
We notice that the dispersion in the D-terms estimated from the visibilities of 
the different sources is large (see the uncertainties in the amplitude averages!).
Such a large dispersion in the D-term estimates
is indicative of a strong coupling (in the LPCAL fitting) between the 
polarization leakage and the polarized source components, which maps into a 
poor modeling of the polarization leakage. We also notice 
that the visibilities in the cross-hand correlations are quite sensitive to the 
leakage, so the final full-polarization GMVA images may differ notably, 
depending on the different schemes used for the estimate of the D-terms.
 
However, it would be expected that 
the main polarization features in the images (i.e., the source components with the 
strongest polarized emission) are rather insensitive to changes in the estimated 
D-terms. Moreover, there may also be correlations in the D-terms (i.e., couplings in
the D-term estimates at the different antennas, resulting from the fitting procedure 
in LPCAL), such that images obtained from the use of different sets of 
D-terms do not differ significantly. We performed a quantitative analysis of how strongly the GMVA polarization 
images differ as a function of the different weighting schemes in the D-term averaging.
In our analysis, we have estimated the highest dynamic range achievable in the 
polarization images, such that the result should be nearly independent of the weights 
applied in the D-term averaging. This analysis is based on a Monte Carlo 
approach, and is described in the following lines. For each source:

\begin{enumerate}

\item We generate the dirty image of the Stokes parameters Q and U, 
calibrated using the vector-averaged D-terms (which are obtained 
as described in Sect. \ref{V}). Let us call this result the {\em reference 
polarization image}.

\item We compute a new vector-average of the D-terms, but using random weights 
for the different sources (weights uniformly distributed between 0 and 1).

\item We generate the dirty images of the Stokes parameters Q and U using these 
new D-terms, and subtract these images from the reference image (i.e., that 
generated in step 1). Let us call these results {\em differential polarization 
images}.

\item We compute the ratio between the intensity peaks in the differential 
images (i.e., those generated in step 3) and the intensity peak in the 
reference image (i.e., that generated in step 1).

\item We iterate steps 2 to 4.

\end{enumerate}

The intensity peaks in the differential images (i.e., those in step 3) give us 
an estimate of how different are the images when we (randomly) change the weights of 
the D-terms in the average. Ideally, the peaks in these images should be zero (i.e., the images 
generated in step 1 and step 3 should be equal), regardless of the weighting 
scheme used to compute the average of the D-terms. Hence, the 
ratio between the intensity peak in the differential images and the peak of the (D-term 
corrected) polarization image will be a measure of the dynamic range 
achievable, such that the images are independent of the different weights applied to the 
D-terms. In other words, the peaks in the differential images are 
lower bounds to the flux density per unit beam of the source components that are 
almost insensitive to changes in the D-terms averaging.

\begin{figure}
\centering
\includegraphics[width=9cm,angle=0]{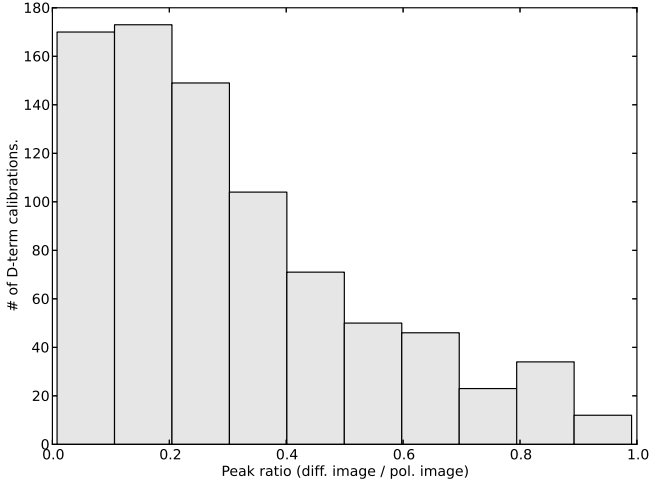}
\caption{Distribution of the intensity peaks of the {\em differential polarization 
images} (see text) of source 3C\,345, divided the peak of the reference polarization 
image.}
\label{Dterms}
\end{figure}

In Fig. \ref{Dterms}, we show the distribution of intensity peaks in the differential
polarization images of source 3C\,345, using a total of 1000 Monte Carlo iterations. 
The cut-off probability of 95\% (i.e., 2 sigma) for the null hypothesis of a 
false detection corresponds to a peak intensity of $\sim$0.65 times the peak 
in the polarization image. Hence, any 
source component with a flux density larger than $\sim$0.65 times that of the peak can
be considered as real, with a confidence of 95\%. 

\begin{figure}
\centering
\includegraphics[width=9cm,angle=0]{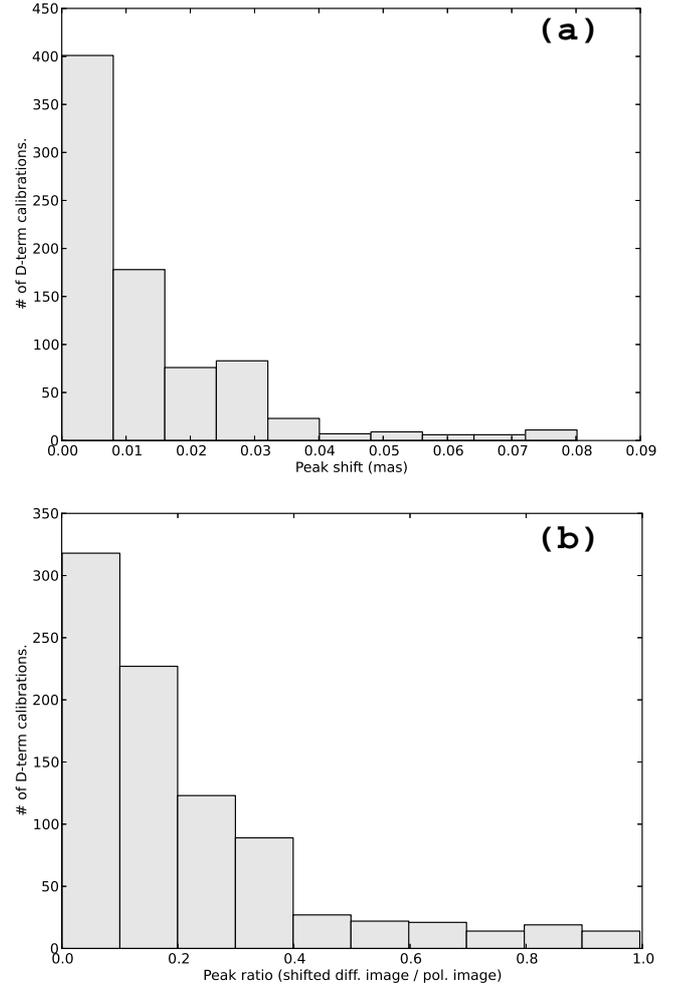}
\caption{(a) Distribution of the shifts in the intensity peaks of all the 
polarization images obtained in our Monte Carlo D-terms analysis. (b) Same as 
Fig. \ref{Dterms}, but taking into account the peak shifts before computing the 
differential polarization images.}
\label{HistoShift}
\end{figure}

The images obtained using
different D-terms do not only differ in the strength of the polarized features,
but also in their location. Indeed, many of the Monte Carlo iterations where we found 
large peaks in the differential images correspond to cases where the peaks of 
the Monte Carlo images were slightly shifted with respect to the peak of the 
reference polarization image. We show in Fig. 
\ref{HistoShift}(a) the distribution of shifts between the peak of the reference image 
and the peak of the images obtained from all the Monte Carlo iterations. 
Most of the Monte Carlo images have their peaks at less than 30\,$\mu$as away from the
peak of our reference polarization image (this is roughly the size of the minor axis 
of our beam). If we take into account these small shifts in the computation of the 
differential polarization images, the resulting peaks of the new differential images 
are quite lower than those without shifting, as we show in Fig. \ref{HistoShift}(b). 
Hence, if the small shifts are corrected, the final images 
do not typically differ at a level of more than 50--60\% of the source 
peak (with a confidence interval of 95\%). The position shifts obtained from the 
different D-term calibration also imply that the astrometric precision in the
location of the polarized emission is of the order of $\sim$30\,$\mu$as 
(roughly the size of the minor axis of the synthesized beam).

\section{Representative images}
\label{VI}

\begin{figure}
\centering
\includegraphics[width=9cm,angle=0]{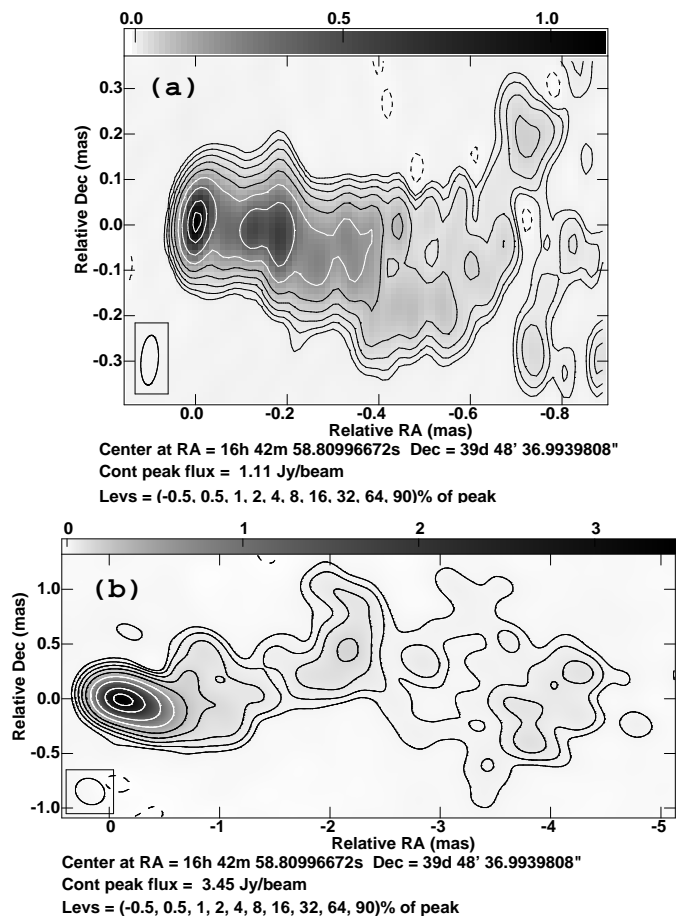}
\caption{Total-intensity images of 3C\,345 obtained from the analysis of the GMVA 
data taken on May 2010. The full width at half maximum (FWHM) of the restoring beams 
are shown at the bottom-left corners. (a) using uniform weighting of the visibilities
(restoring beam with FWHM of 110$\times$38\,$\mu$as with a position angle of $-4.37$\,deg.). 
(b) using natural weighting of the visibilities and tapering longest baselines (to enhance 
the sensitivity to extended structures; FWHM of 270$\times$230\,$\mu$as with a position 
angle of 62\,deg.). }
\label{Image}
\end{figure}

\begin{figure*}
\centering
\includegraphics[width=18cm,angle=0]{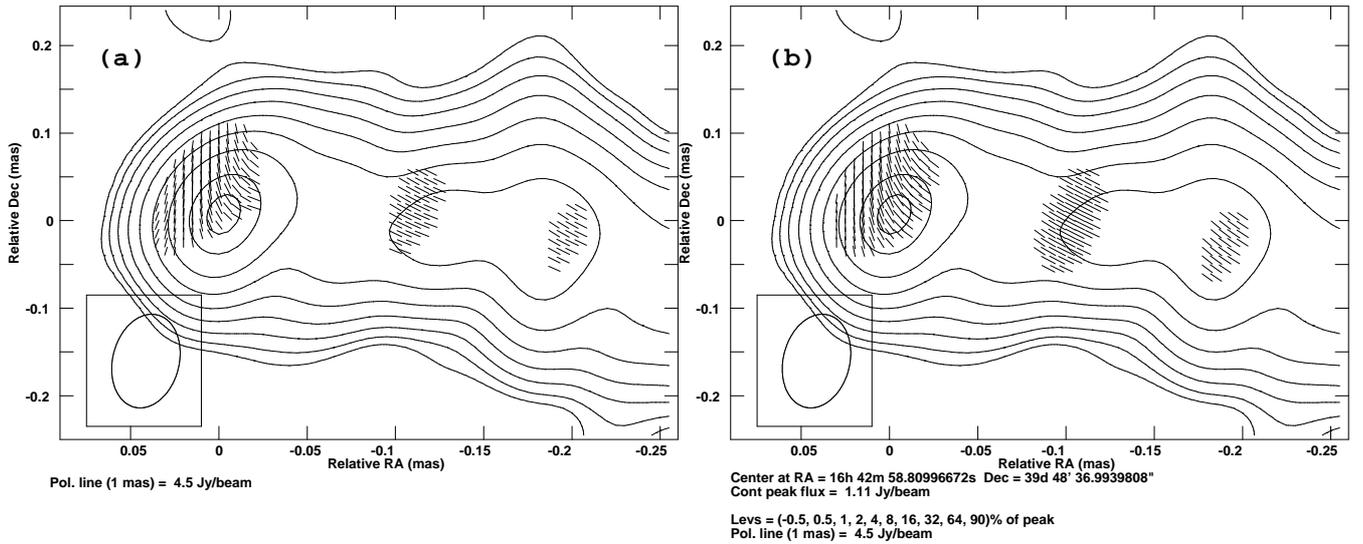}
\caption{Polarization images (superimposed to the total-intensity image) of 3C\,345. 
The FWHM of the restoring beam is shown at the bottom-left corners. (a) averaging the 
D-terms estimated from the visibilities of 3C\,345, BLLAC, and 0716+714. (b) using 
the D-terms directly estimated from the visibilities of 3C\,345. }
\label{Image2}
\end{figure*}

\begin{figure}
\centering
\includegraphics[width=5cm,angle=270]{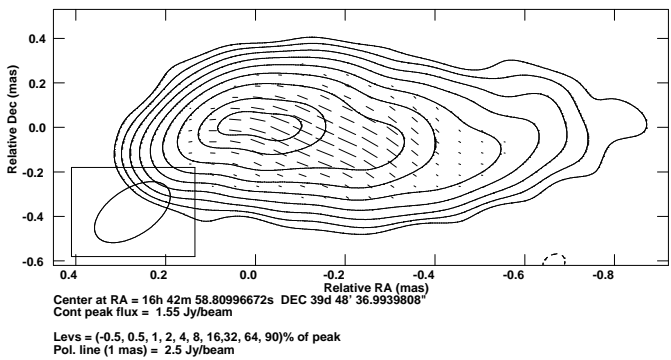}
\caption{VLBI image of the inner core region of 3C345 at 43 GHz (Jorstad et al., in prep.), 
observed on 19 May 2010.}
\label{P43GHz}
\end{figure}

Once the data are calibrated as described in the previous sections, they are ready for a full-polarization 
imaging (taking into account the polarization limitations described in Sect. \ref{Fidelity}). We present, 
in Fig. \ref{Image}, a sample image of the source 3C\,345
obtained from the GMVA observations reported here. The high quality of the GMVA data allows us to recover 
extended jet structure distant from the core, after careful imaging, including iterative amplitude
self-calibration and uv-tapering.
We also show in Fig. \ref{Image2} two polarization images of the same source, obtained 
from different estimates of the antenna D-terms (i.e., averaging the D-terms estimated from 
the visibilities of a selection of sources or using the D-terms just estimated from the 
visibilities 3C\,345). The polarization is very similar in both 
images (we applied a cut-off at 60\% of the polarization peak). There is polarized emission at 
the north-east side of the core, where the electric-vector 
position angle (EVPA) is perpendicular to the jet. Then, the electric vector position angle rotates as the distance to the core 
increases westwards. The polarization images in Fig. \ref{Image2} can be compared to another image 
obtained from VLBI observations at 43\,GHz (Jorstad et al., in prep.) taken on 19 May 2010, 
only a few days before our GMVA session. We show the 43\,GHz image (only the part 
near the VLBI core of the source) in Fig. \ref{P43GHz}. The electric vector position angle at 43\,GHz is very similar to 
that of the optically-thin components in Fig. \ref{Image2} (i.e., the western 
components, away from the core at 86\,GHz), although we notice that the absolute electric-vector position angle (i.e., a 
possible global R-L phase offset at the reference antenna, which would map into a global 
rotation of all the polarization vectors in the image) has not been determined in our observations. 
We also notice that the polarized core component with north-south 
electric vector position angle at 86\,GHz is not detected at 43\,GHz. Possible reasons of this discrepancy in the polarized emission 
at different frequencies could be opacity, Faraday rotation, or blending (due to the larger beam at 
43\,GHz). 
A deep analysis of Figs. \ref{Image2} and \ref{P43GHz} lies beyond the scope of the 
objective of this paper, and will be published elsewhere.

\section{Summary}
\label{VII}

We report a well-defined calibration pipeline for global millimeter-VLBI (GMVA) observations. 
With this pipeline, it is possible to estimate all the instrumental effects in 
an optimum way, dealing with the particulars of the (typically complicated) schedules 
of global 3mm VLBI observations and the inherent complications due to the high observing frequency (86\,GHz).
All the scripts used in the pipeline are written in a generic way, so they can be easily executed
and adapted for all the GMVA (and eventually non-GMVA) datasets. Indeed, these scripts will still 
be valid if new stations eventually join the GMVA in a near future. 

The scripts allow us to perform manual phase calibration (i.e., alignment of the phases among the different
sub-bands) regardless of the subarraying conditions typically found in the data. 
The script also corrects for time dependent phase and delay drifts between subbands caused
by variations in the electronics of the antenna receivers.

We perform the global fringe fitting (GFF) by optimizing the integration time of the fringes within 
the real coherence time of the visibilities. In the case of the GMVA, we show that at 86 GHz
integration times of up to several minutes maximize the SNR of the fringes, indicative of an only
moderate atmospheric degradation of the incoming phase.

The visibility amplitudes are calibrated by fitting the temperature of the receivers of each antenna 
(and polarization) to the distribution of system temperatures over airmass. The opacity is then 
directly derived from the ambient and system temperatures for each antenna and time. For the cases 
of antennas where the atmospheric absorption is directly accounted in the amplitude calibration, we 
can still refine the opacity correction with our approach.

For the polarization calibration, we perform manual phase calibration on the cross-polarization 
visibilities (i.e., we align the phases of the cross-polarization visibilities among the sub-bands)
by imposing the same fringe rates in both polarizations for all the antennas and times.
This calibration allows us to determine the leakage in the receivers (i.e., the D-terms) using 
the data of all sub-bands together, thus duplicating the SNR with respect to the D-terms estimated 
from independent fits to the different subbands.

Our scripts also allow us to perform a Monte Carlo analysis to determine the effect of different
D-term calibrations on the final full-polarization images. For the data reported here, a cut-off 
in the polarization 
images at a level of 50--60\% of the intensity peak generates images that are typically very similar, 
regardless of the D-term calibration. The absolute position of the polarization features can also 
be affected by the D-term calibration, but this effect is not larger than the size of the the 
synthesized beam.

As an example of our pipeline output, we show full-polarization images of source 3C\,345, obtained from
observations performed during the GMVA session of May 2010. The polarization images show a strong 
component at the northeast side of the source core, with a north-south electric vector position angle that rotates counter-clockwise 
along the jet (i.e., along the west direction). There are also two polarization components at about
0.1 and 0.2\,mas from the core, with an electric vector position angle aligned to the direction of the jet and similar to 
the electric vector position angle observed at 43\,GHz from VLBA observations taken a few days before our GMVA session.

In forthcoming GMVA sessions, we expect to be able to provide the end user with fully calibrated 
datasets, obtained short after the data correlation.

\begin{acknowledgements}

The National Radio Astronomy
Observatory is a facility of the National Science Foundation
operated under cooperative agreement by Associated Universities,
Inc. Based on observations with the 100-m telescope of the MPIfR 
(Max-Planck-Institut f\"ur Radioastronomie) at Effelsberg.
Based on observations carried out with the IRAM Plateau de Bure 
Interferometer and the IRAM telescope at Pico Veleta. IRAM is 
supported by INSU/CNRS (France), MPG (Germany) and IGN (Spain).

\end{acknowledgements}

\end{document}